\documentclass[11pt,a4paper]{article}
\usepackage[hyperref]{emnlp-ijcnlp-2019}
\usepackage{times}
\usepackage{latexsym}
\usepackage{amsmath, amssymb}

\usepackage[colorinlistoftodos]{todonotes}

\usepackage{booktabs}
\newcommand{\ra}[1]{\renewcommand{\arraystretch}{#1}}

\usepackage{url}

\aclfinalcopy 

\title{Modelling Stopping Criteria for Search Results using Poisson Processes}

\author{Alison Sneyd \and Mark Stevenson\\
  Natural Language Processing Group\\
  Department of Computer Science\\
   University of Sheffield \\
  {\tt a.sneyd,mark.stevenson@sheffield.ac.uk} \\}

\date{}

\begin{document}
\maketitle
\begin{abstract}
Text retrieval systems often return large sets of  documents, particularly when applied to large collections. Stopping criteria can reduce the number of these documents that need to be manually evaluated for relevance by predicting when a suitable level of recall has been achieved. In this work, a novel method for determining a stopping criterion is proposed that models the rate at which relevant documents occur using a Poisson process. This method allows a user to specify both a minimum desired level of recall to achieve and a desired probability of having achieved it. We evaluate our method on a public dataset and compare it with previous techniques for determining stopping criteria.
\end{abstract}

\section{Introduction}\label{sec:intro}
Document retrieval systems are widely used in information retrieval (IR) to return a set of documents in response to a user query. They often return large numbers of documents, particularly for large collections, which places a significant burden on a user who has to manually review them to determine which are relevant (i.e. of interest). The problem is particularly acute in scenarios where relevant documents are rare in the set of retrieved documents, or where there is a high cost associated with missing a relevant document. 

Systematic reviews of medical literature are a good example of such a task. They aim to identify and summarise all  available evidence for a specific research question such as \lq Is systemic inflammation present in stable chronic obstructive pulmonary disease?\rq \cite{gan2004association}. Evidence included in systematic reviews is typically identified by executing a Boolean query optimised for high recall, with the results manually reviewed by domain experts. However, this process is extremely expensive and a single review may cost up to a quarter of a million dollars and take over a year to complete \citep{McGowan, Pohl}. 
%
%
A range of studies have demonstrated that ranking the results of the Boolean query can support the identification of relevant studies by placing them higher in the ranking, e.g. \cite{Miwa2014,Wallace2010, OMara-Eves2015, Shemilt}. However, ranking alone does not reduce the manual review workload since it does not inform reviewers about when they can stop looking at search results. 

Search stopping criteria are methods for determining when manual examination of search results can stop. They essentially provide information about the risk of missing relevant documents that is associated with not examining some portion of the documents returned by a query. 
Stopping criteria have been proposed for technology assisted reviews \citep{C&G} and IR test collection development \citep{Losada}. 

\citet{C&G} introduced the target and knee stopping methods. The former  involves examining randomly chosen documents until a specified number of relevant documents have been found. The number of relevant documents required is determined using statistical inference and guarantees a minimum recall score with a specified probability. The knee method employs a knee detection algorithm \citep{knee} to estimate when the gain curve of recall versus effort has exceeded a certain tipping point, called the slope ratio. \citet{C&G} found that this method was better able to predict the most efficient stopping point than the target method, but did not provide any probabilistic guarantees about the level of recall that would be achieved.   
Additionally, \citet{Losada} proposed simple baselines and used power curves to determine stopping criteria.  A significant limitation of previous approaches is that they are evaluated using a single ranking algorithm, but for a stopping approach to be generally useful it must be reliable for a range of ranking algorithms.

This paper proposes a novel method for generating search stopping criteria based on an inhomogeneous Poisson process.
The method assumes the rate at which relevant documents occur in the ranking decreases exponentially and  predicts how many relevant documents need to be found in order to achieve a defined level of recall. 
This approach, along with the knee and target methods, is evaluated using data from a shared task 
for which over 30 submissions are available. This allows approaches to be evaluated against a wide range of rankings of varying effectiveness. We found that, while each method returns a reliably high level of recall, the proposed method produces the most accurate predictions while also being robust to ranking quality.\footnote{Code available at: \url{https://github.com/alisonsneyd/poisson\_stopping\_method}}


\section{Method}\label{sec:method}
Our proposed method models the rate at which relevant documents occur in a ranked set of retrieved documents, using an exponential function learned from a sample of the data. It is then possible to predict the total number of relevant documents to a desired probability using a Poisson process, and from this, to predict the rank at which a desired level of recall is achieved. 

\subsection{Poisson Process}

Poisson processes model the occurrence of random events that take place at a given rate \cite{Poisson}. 
It is assumed that the events are independent and that the number of occurrences in an interval follows a Poisson distribution. 
%
An inhomogeneous (or non-homogeneous) Poisson process has a variable rate function, $\lambda$, where the rate at which events occur is a function of the space over which the process is defined. 
Let:
\begin{equation}\label{eqnLam}
\Lambda(a,b) = \int_a^b \lambda(x)dx,
\end{equation}
and let
$N(a,b)$ denote the number of events  occurring in the interval  $(a,b]$. Then:
\begin{equation}\label{eqnprob}
	   P\left(N(a, b) = r\right) = \frac{[\Lambda(a,b)]^r}{r!} e^{-\Lambda(a,b)}.
\end{equation}
In particular, $N$ is a Poisson random variable with expected value $\Lambda(a,b)$.

We model the occurrence of relevant documents in a set of ranked, retrieved documents as an inhomogeneous Poisson process, where 
 $\lambda$ is an exponential function (i.e. $\lambda(x) = de^{kx}$). So, from Equation~\ref{eqnLam}: 
\begin{align*}\label{eqnLamExp}  \Lambda(a,b) & = 
	    \int_a^b de^{kx} dx \\ & = \frac{d}{k}\left(e^{kb} - e^{ka} \right).
\end{align*}
 We are interested in knowing the total number of relevant documents and so restrict our attention to the interval $(0,n]$, where $n$ is the number of documents in the set, yielding:
\begin{equation}\label{eqnLamInt}
	   \Lambda(0, n) = 
	   \frac{d}{k}(e^{kn} - 1).
\end{equation}
By Equations~\ref{eqnprob} and \ref{eqnLamInt}, the probability of there being $r$ relevant documents in the document set is:
\begin{equation*}
	   P\left(N(0,n) = r\right) = 
	   \frac{\left[\frac{d}{k}(e^{kn} -1)\right]^r}{r!} e^{-\left(\frac{d}{k}\left(e^{kn} - 1\right)\right)}.
\end{equation*}

Since $N(0,n)$ is a Poisson random variable, the upper limit of the number of relevant documents in the collection ($R$), subject to a given probability, can be estimated using the cumulative distribution function of $N(0,n)$. The number of relevant documents required to achieve a desired level of recall ($T$) is then $RT$. The proposed stopping criteria is to cease relevance judgments when $ \left \lceil{RT}\right \rceil$ relevant documents have been found. This method is flexible and can be adapted to whatever level of recall a user desires, and to the level of certainty they want associated with it.

\subsection{Rate Estimation}

The approach proceeds by examining the first $\alpha$ documents in the ranking, and then predicting the total number of relevant documents in the entire ranking. If the number of relevant documents observed in the first $\alpha$ documents ($rel(\alpha)$) is greater than or equal to the number of relevant documents needed to achieve the desired level of recall (i.e. $ rel(\alpha) \geq \left \lceil{RT}\right \rceil$), the process stops and no more documents are examined. Otherwise, the next $\beta$ documents are added to the pool of examined documents and the process repeated.

Accurate estimation of the rate at which relevant documents occur in the ranking is an important component of the proposed approach. The parameters of the rate function are estimated as follows: Split the documents which have been examined so far into $m$ sub-intervals and define a set of points $(x_i,y_i)$, $i = 1, \ldots, m$, where $x_i$ is the midpoint of the $i^{th}$ interval and $y_i$ is the number of relevant documents it contains. 
Fit a curve of the form $de^{kx}$, $d,k \in \mathbb{R}$, to the points. 

This approach assumes that relevant documents occur frequently enough for their rate to be predicted and that this rate decreases approximately exponentially, thereby implying the shape of the gain curve (number of relevant documents versus rank) will be roughly convex. While these assumptions frequently hold, they are not guaranteed to do so and they are therefore tested in multiple ways. 

Firstly, if fewer than a given threshold, $\gamma$, of relevant documents are found in the initial sample of $\alpha$ documents, it is assumed that either relevant documents are too rare for accurate estimates or that the ranking places them later, meaning their rate is not exponential. No stopping point is predicted and all documents in the ranking are examined (while this is costly in terms of document judgment, it preserves recall and is the default approach in systematic reviews).


Secondly, the Poisson process  
is dependent on $\lambda(x)$ reliably estimating the rate at which relevant documents occur. Therefore when a curve is fitted, its accuracy is tested as follows: Suppose the sample ends at rank $n$. The number of relevant documents estimated by $\lambda(x)$ at rank $n$, given by
$\sum_{i=1}^n \lambda(i)$,
is compared to the actual number of relevant documents in the sample ($rel(n)$). If the number of relevant documents found is less than a scalar multiple ($\delta$) of the predicted number (i.e. if $rel(n) < \delta \times \sum_{i=1}^n \lambda(i)$) then the curve is assumed to be a bad predictor. No stopping point is predicted and the next $\beta$ documents in the ranking are examined. 


\section{Experiments}
\subsection{Data} 
The dataset used for the experiments originated from the CLEF 2017 e-Health Lab Task 2 ``Technology Assisted Reviews in Empirical Medicine" \citep{CLEF17-1}.
The task required participants to identify relevant studies for a set of 30 Diagnostic Test Accuracy systematic reviews (known as ``topics'') from the Cochrane Library.\footnote{\url{https://www.cochranelibrary.com/}} 
Participants were provided with the (unordered) set of documents retrieved by a complex Boolean query and asked to rank them based on relevance to the review topic. 

The number of documents retrieved by the Boolean queries for the 30 topics ranged from 64 to 12,807, with median 2,070, while the total number of documents across all topics was 117,562. The number of relevant documents ranged from 2 to 460, with median 38. On average, only 1.58\% of documents were relevant across the 30 topics.\footnote{The CLEF data includes relevance judgments at two levels: abstract and content. The abstract level information is used for the experiments described in this paper.}

A total of 33 runs ranking the entire set of articles  were submitted by 11 groups of participants.\footnote{Available at  \url{https://github.com/CLEF-TAR}} The rankings produced by these runs were based on a wide range of techniques, including manual judgments of document relevance, and automated techniques based on (semi-)supervised classifiers and relevance feedback. These runs contain a diverse set of rankings of varying effectiveness, making them very suitable for testing the robustness of search stopping criteria.



\subsection{Evaluation Metrics}
The goal is to reduce the amount of manual judgment required, subject to maintaining a minimum level of recall.
Following \citet{C&G}, the following metrics were used: For a stopping criteria $S$ and topic $T$, 
%
%
the \textit{effort} of $S$ in $T$ is defined as $|E_T|$, where $E_T$ is the set of documents that were examined before stopping.
For a desired level of recall $R$, the \textit{acceptability} of $S$ with respect to $T$ is defined as:

\vspace{-3eX}
\begin{equation*}
  acceptability(S_T)=\begin{cases}
    1, & \text{if $recall(S_T) \geq R$},\\
    0, & \text{else}.
  \end{cases}
\end{equation*}

As sets of runs of topics are being evaluated, additional metrics are used. For a stopping criteria $S$ considered over a set of topics in a run $\mathcal{R}$,  the \textit{reliability} of $S$ with respect to $\mathcal{R}$ is:
\begin{equation*}
\frac{|\{T \in \mathcal{R} : acceptability(S_T) = 1\}|}{|\mathcal{R}|}.
\end{equation*}
$S$ is considered \textit{reliable} over a set of runs if its mean reliability over the runs is not less than a desired threshold $p$, which is equivalent to the proportion of all topics in all runs that are acceptable being greater than or equal to $p$. 
The \textit{effort} of $S$ with respect to  $\mathcal{R}$ is $\sum_{T \in \mathcal{R}} |E_T|$. The \textit{percentage of effort saved} by $S$ over $R$ will denote the percentage of all  documents in the set that do not have to be examined: $$100 \times \sum_{T \in \mathcal{R}}\frac{|T| - |E_T|}{|T|}.$$
%

%


\subsection{Approaches}
The Poisson process approach (Section \ref{sec:method}) and knee and target methods (Section \ref{sec:intro}) were implemented. For all methods, the minimum level of recall was set to 0.7 and the reliability threshold to 0.95, following \citet{C&G}. 

For the Poisson process ({\bf PP}), the initial sample size ($\alpha$) and increases in batches ($\beta$) were set to $0.3 |T|$ and $0.05|T|$ respectively.
The value of $\alpha$ was chosen as it is large enough to predict the rate curve with reasonable reliability but without requiring too many documents to be examined. The $\gamma$ and $\delta$ parameters were set to 20 and the $0.7$ respectively. 


For the target method ({\bf TM)}, we follow \citet{C&G} in requiring that a set of 10 relevant documents be identified before the stopping point is predicted since they demonstrated that this produced a minimum recall of 0.7 with 0.95 probability. 

The knee method requires a slope ratio to be set given by $\epsilon + 6 - min(relret, \epsilon)$, where $relret$ is the number of relevant documents retrieved for a given rank and $\epsilon$ is a variable that prevents the method stopping too soon. A version of the approach ({\bf KM-default}) was implemented using $\epsilon = 150$, as recommended by \citet{C&G}. 
The value of $\epsilon$ was also tuned using 3-fold cross-validation with $\epsilon \in \{0, 25, 50, 100, 150, 200\}$. The smallest reliable value for $\epsilon$ was found to be 50 in each fold and consequently a second version  ({\bf KM-tuned}) was implemented using this value.


For comparison, an Oracle ({\bf OR}) method was also implemented which stops at the point in the ranking where 70\% recall has been achieved.

\section{Results}

Each method was found to be reliable (i.e. achieved the desired recall of 0.7 for at least 95\% of topics). The top part of Table \ref{table17_all} shows the results over all topics in all 33 runs. 
The proposed approach, PP, required less effort than other methods (TM, KM-tuned and KM-default). PP allows the desired recall to be achieved without examining 42.1\% of the documents, thereby allowing 4.9\% more of the documents to go unexamined than the next most efficient approach (KM-tuned). 

Further analysis was carried out to explore the effect of the quality of the initial ranking on the performance of the various approaches. The 33 runs were ranked based on their Area Under Recall Curve (AURC) scores in the official evaluation \citep{CLEF17-1}. (The AURC metric computes the area under the cumulative recall curve normalized by the optimal area.) The top, middle and bottom five runs in the rankings were selected. The top five had AURC scores in the range 0.92 - 0.93, indicating effective rankings in which most of the relevant documents appear early. The bottom five runs had scores in the range 0.48 - 0.60, indicating the rankings are not much different from random orderings of the documents.

PP was the most efficient in terms of effort for the middle and bottom five runs, and the second most efficient for the top five runs. KM-tuned required the least effort for the top five runs, but the second most for the middle five runs and the third most for the bottom five runs. As might be expected, each method required the least effort for the top five runs and the most effort for the bottom five. 

\begin{table}[h!]
\centering
 \ra{1.3}
\begin{tabular}{@{}rrr@{}}
\toprule
& & Mean $\%$\\
& Mean Eff.  & Eff. Saved \\
\hline
\multicolumn{3}{c}{All 33 runs}\\
\hline
TM         &   79,111 &       32.7\% \\
KM-default &  102,681 &       12.7\% \\
KM-tuned   &   73,826 &       37.2\% \\
PP         &   \textbf{68,122} &       \textbf{42.1\%} \\
OR         &   33,760 &       71.3\% \\
\hline
\multicolumn{3}{c}{Top 5 runs}\\
\hline
TM         &  56,762 &     51.7\% \\
KM-default &  87,886 &     25.2\% \\
KM-tuned   &  \textbf{33,841} &     \textbf{71.2\%} \\
PP         &  54,792 &     53.4\% \\
OR         &   7,202 &     93.9\% \\
\hline
\multicolumn{3}{c}{Middle 5 runs}\\
\hline
TM         &   76,704 &      34.8\% \\
KM-default &  107,703 &       8.4\% \\
KM-tuned   &   75,888 &      35.4\% \\
PP         &   \textbf{60,311} &      \textbf{48.7\%} \\
OR         &   27,565 &      76.6\% \\
\hline
\multicolumn{3}{c}{Bottom 5 runs}\\
\hline
TM         &    104,620 &       11.0\% \\
KM-default &    111,055 &        5.5\% \\
KM-tuned   &    110,778 &        5.8\% \\
PP         &    \textbf{100,622} &       \textbf{14.4\%} \\
OR         &     74,389 &       36.7\% \\
\bottomrule
\end{tabular}
\caption{Mean effort and mean percentage of effort saved for the 33 CLEF 2017 runs. 
Results are shown for all runs plus the top, middle and bottom five runs. Highlighted figures are the best non-Oracle results.}
\label{table17_all}
\end{table}

\begin{figure}[h]
\centering
\includegraphics[width = 7.5cm]{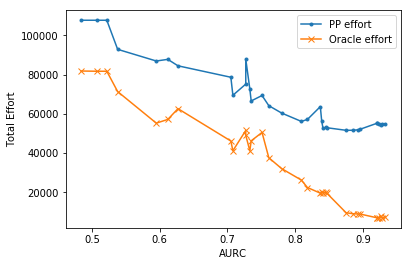}
\caption{Oracle and Poisson process effort versus AURC score.}
\label{oracle_effort}
\end{figure}

A graph of the OR and PP effort scores for each run is shown in Figure~\ref{oracle_effort}.

The difference in performance between the tuned and untuned version of the knee method (KM-tuned and KM-default) demonstrates the sensitivity of this method to appropriate setting of parameter values. These results demonstrate that the approach performs strongly for effective rankings, such as the one it was developed for  \citep{C&GCAL}, but is not robust to less effective ones. The PP approach is less sensitive to ranking quality (which is unlikely to be known a priori).

\begin{figure}[h]
\centering
\includegraphics[width = 7.5cm]{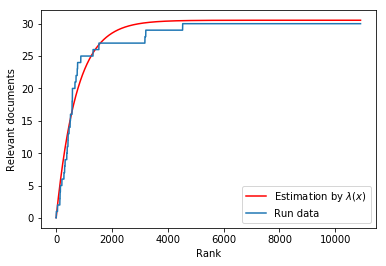}
\includegraphics[width = 7.5cm]{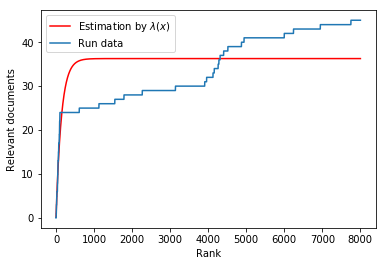}
\caption{The gain curves of two rankings from CLEF 2017 runs, compared to the gain curve estimated by $\lambda$.}
\label{curveex}
\end{figure}

Two examples of the actual gain curves of a ranking and the gain curve estimated by the rate function $\lambda(x)$ learned by the PP approach are shown in Figure \ref{curveex}. In the top graph, the rate at which relevant documents occur decreases exponentially and the high similarity between the two curves shows $\lambda$ is fit well. In the bottom graph, the rate at which relevant documents occur does not decrease exponentially. However, PP still returns the desired level of recall because the probability of the Poisson process was set to a high value (0.95).

\section{Conclusion}

This work proposed a novel method for determining a stopping point in search results using a Poisson process. The method was applied to 33 varied CLEF 2017 runs and proved reliable in terms of recall, while also requiring substantially less effort and demonstrating less sensitivity to ranking quality than existing methods. On average, the Poisson process saved manual judgment of over $40\%$ of the documents associated with a run.
Although our method has been explored in the context of systematic reviews, it could be applied to a variety of other document retrieval problems.

\bibliography{emnlp-ijcnlp-2019}
\bibliographystyle{acl_natbib}

\end{document}